\documentclass[aps,pre,reprint]{revtex4-2}
\usepackage{graphicx, amsmath, braket, color, soul, graphicx, amsfonts, appendix, comment} 

\usepackage[dvipsnames]{xcolor}
\usepackage{chngcntr}
\usepackage{float}
\usepackage[hidelinks,colorlinks=true,linkcolor=green,citecolor=green]{hyperref}
\usepackage{siunitx}

\begin{document}

\title{Laddering of a knitted fabric: a topology-induced failure}

\author{Antoine Faulconnier$^a$}

\author{Laura Michel$^b$}
  
\author{Mokhtar Adda-Bedia$^a$}

\author{J\'er\^ome Crassous$^{b,c}$}
   
\author{Audrey Steinberger$^a$}
\email{audrey.steinberger@ens-lyon.fr}

\affiliation{$^a$CNRS, ENS de Lyon, LPENSL, UMR5672, 69342, Lyon cedex 07, France}

\affiliation{$^b$PMMH, CNRS, ESPCI Paris, Universit\'e PSL, Sorbonne Universit\'e, Universit\'e de Paris, 75005, Paris, France}

\affiliation{$^c$Université de Rennes, 35000, Rennes, France}

\begin{abstract}
Laddering is the propagation of a topological defect in an everyday-life material: weft knitted fabrics, following a broken thread or a dropped stitch. What is a minor frustration when damaging a pair of tights is a more serious issue for industrial-scale production, but might inspire new solutions to limit and mitigate damage to architected materials.
In this work, laddering is investigated in a pre-stressed model knit through experiments and Discrete Element Rod simulations. The control parameter is the initial tension applied on the fabric. A force threshold due to the stitch’s natural curvature is evidenced. It controls both the propagation onset and arrest, as tension is relaxed by the thread length freed by ladder growth, and enables damage prediction at moderate tension. Furthermore, we uncovered that the laddering velocity is of the order of the velocity of bending waves and exhibits an unexpected linear scaling with the fabric tension, that arises from a complex combination of elastic and friction forces.  Finally, we discuss the implications of our results from the perspective of damage control and mitigation.
\end{abstract}

\maketitle

There is a growing interest of creating architected materials with tunable mechanical response and controlled failure properties~ \cite{Shenhav2019,Gao2024,deWaal_PRR2025,deWaal_PRL2025}. Knitted fabrics constitute an archetype of mechanical metamaterials where both the mechanical properties of the constituents and the underlying topology and geometry come into play. Unlike conventional architected materials where damage is mediated by cracks, laddering is a mode of failure specific to knitted fabrics, owing to their specific topology~\cite{Shimamoto2025,poincloux2018geometry}. Laddering creates structures that resemble dislocations and the material does not fail completely, thus paving the way to a profitable exploitation of such specific damage for engineering applications.

A weft knitted fabric is a network of interlaced loops, called stitches, built from a single yarn and stabilized by its topology. As mechanical metamaterials, their geometry can be tuned to obtain tailored 3D-structures~\cite{Niu2025,Tajiri2025} and specific mechanical properties~\cite{Singal2024,duPasquier2025,Matsumoto2025} for a variety of applications, such as biomedical products~\cite{Li2013,Zhang2018,Singal2024,duPasquier2025}, wearable sensors and actuators~\cite{Abel2013,Ahmed2025,Mahadevan2026,Abbas2026}, soft robotics \cite{Elmoughni2021,Luo2022,Yang2023,Sanchez2023,duPasquier2024}, anti-ballistic protection~\cite{Zhou2025}, and knitted composite preforms for aeronautical applications~\cite{Marvin1961,Savci2000,Leong2000}.

\begin{figure}[ht]
    \centering
    \includegraphics[width=0.9\linewidth]{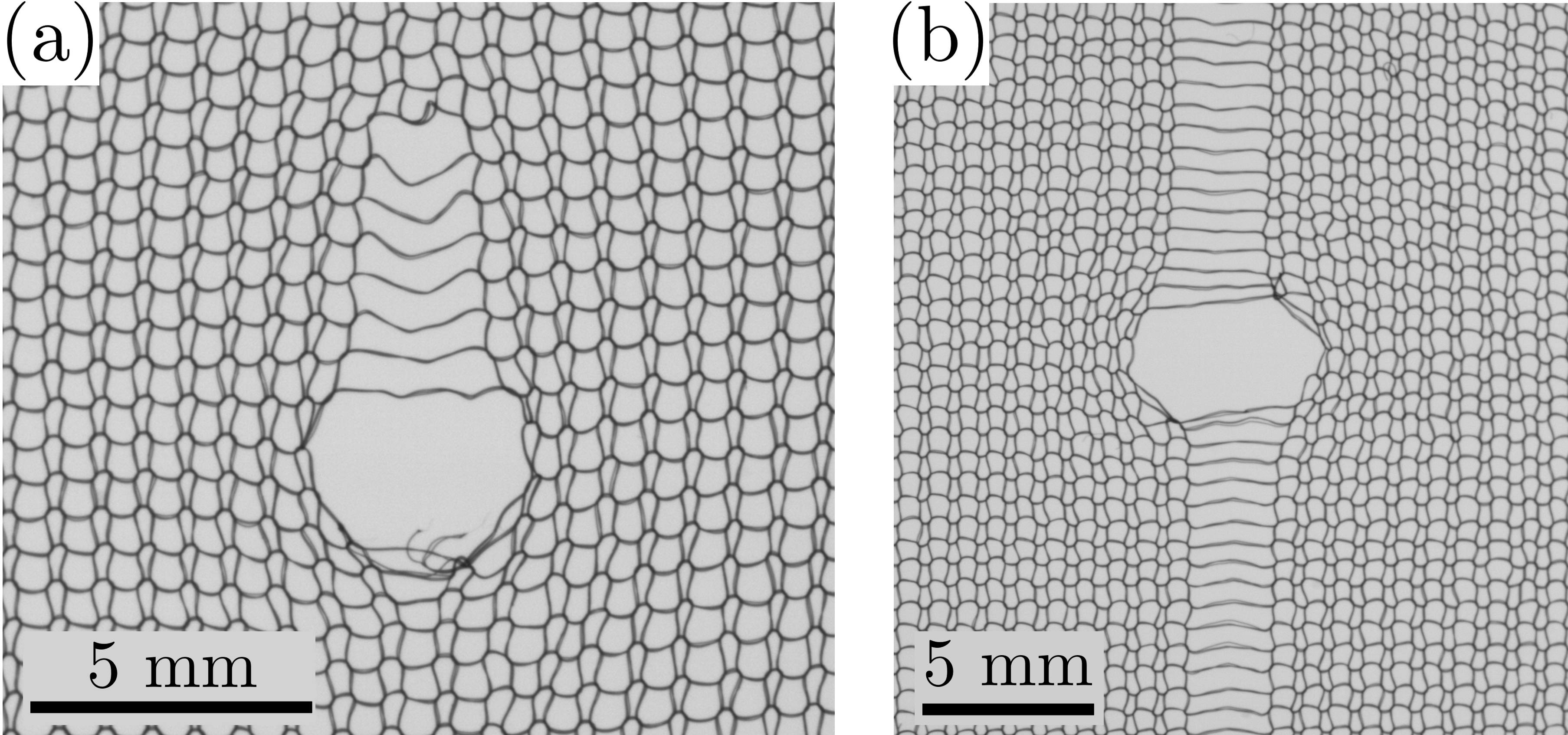}
    \caption{Close-up photographs picturing the final state of a ladder defect in a commercial tight (DIM, Mes Essentiels).}
    \label{fig:fig1}
\end{figure}

As shown in Fig.~\ref{fig:fig1}, a ladder defect is a topological defect in the knitted network arising from a local thread breaking or a dropped stitch during the fabrication process. Consequently, the freed loop can slide through its neighbour in the column (wale) direction, which in turn will be free to slide through its own neighbouring stitch, and so on in a domino effect, forming a growing ladder pattern (see Supplementary Movies and their description in Appendix~\ref{sec:movies}). Such defects are often seen in loose networks such as tights and stockings, sometimes made on purpose for aesthetic effect, but mostly detrimental and leading to the disposal of the fabric. From a mathematical standpoint, laddering is an intrinsic property of weft knitted networks~\cite{Shimamoto2025}.
From a physical perspective, this phenomenon belongs to the field of topological defect propagation~\cite{Mermin1979}, such as the dynamics of dislocations in passive and active nematics~\cite{Kralj2021,Li2025} or in ionic Coulomb crystals \cite{Partner2013} which can be modelled as kink solitons.

Recently, laddering has been addressed as crack a propagation mechanism~\cite{Liu2026} where the focus was on the onset of laddering from pre-cracked knitted samples submitted to lateral strain rate. It was shown that failure through laddering was controlled by a critical stretch that depends mostly on the fabric density and the frictional interactions. In the present letter, we approach laddering from a perspective that does not invoke crack analogy and study both the onset of laddering and its dynamics. We combine experiments and numerical simulations to investigate both the laddering threshold, the final number of laddered stitches and the propagation speed of the defect. We finally discuss our finding from the point of view of damage control and damage mitigation.

\begin{figure}[ht]
    \centering
    \includegraphics[width=0.9\linewidth]{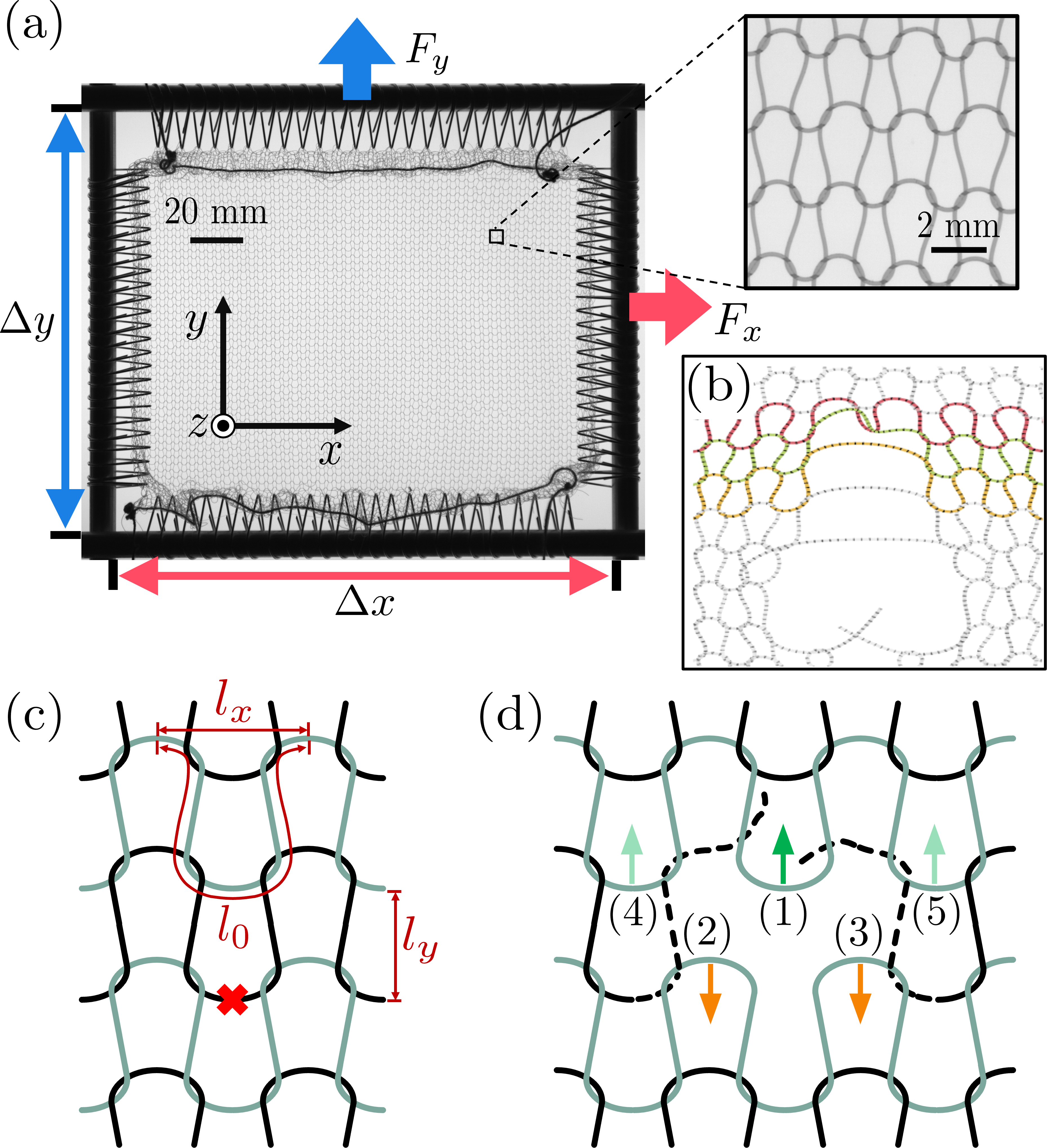}
    \caption{(a) Picture of a knitted sample mounted on the bi-axial tensile machine with a close-up photograph of the sample.
    (b) Snapshot of the simulated knitted fabric while laddering.
    (c) Sketch of the Jersey knit. The red cross indicates where the thread is cut to create a topological defect.
    (d) Sketch showing how laddering is initiated. Colored arrows show the direction of propagation of the ladder that may start from each loop freed by the retraction of the broken thread.
    }
    \label{fig:fig2}
\end{figure}

Samples of model knitted fabrics were made with a single cylindrical Nylon filament (Monofil Madeira) of diameter $d = \SI{0.140}{\milli\meter}$, linear density $\rho_L = \SI{15.8}{\milli\gram\per\meter}$, Young modulus $E = \SI{1.88}{\giga\pascal}$, bending modulus $B = E\pi d^4/64 =\SI{3.55e-8}{\newton\meter\squared}$, and friction coefficient $\mu = 0.4$. 
Each sample is made of $70 \times 70$ stitches knitted in a Jersey pattern using a knitting machine (Silver Reed SK830) to ensure repeatability. Jersey (or stockinette) is the most common pattern where all the stitches are oriented the same way. 
The total length of yarn per stitch  (see Fig.~\ref{fig:fig2}(c)) is $l_0 = 9.74\pm0.02$ mm, corresponding to a loose knit density $d/l_0 = 1/70$, similar to that of commercial tights. 

The central area of the knit sample is mounted with sliding metallic paper-clips on the frame of a bi-axial tensile machine \cite{Crassous2024}, as shown in  Fig.~\ref{fig:fig2}(a). The final sample size is $n_x\times n_y = 50 \times 50$, where $n_x$ is the number of stitches in the $x$ (course) direction and $n_y$ in the $y$ (wale) direction.
The spacings $\Delta x$ and $\Delta y$ between the bars are imposed by stepper motors and measured using magneto-strictive sensors. The applied forces $F_x$ and $F_y$ are measured using strain gauge sensors. 
The sample is first brought to a target spacing $\Delta y$ and $\Delta x$, and a gentle stroke is applied on the knit in order to relax the frictional contacts as much as possible (protocol details are given in Appendix~\ref{sec:StretchingProctocols}). 

To create a topological defect, the thread is locally cut by burning it with the tip of a soldering iron, while filming with a with a high-speed camera (Phantom v2511).
For repeatability, the defect is systematically induced by burning the loop oriented toward $-y$ as indicated by a red cross in Fig.~\ref{fig:fig2}(c). As illustrated in Fig.~\ref{fig:fig2}(d), the cut thread (dashed curve) first releases loop (1), which leads to a ladder propagation in the upward ($+y$) direction. Then, if the tension is high enough, the thread retracts further and frees loop (2) and/or (3), which initiates a downward ($-y$) propagating ladder. Eventually, loop (4) or (5) may also be released, leading to a second upward propagation.This discussion highlights that the knit topology entirely controls the direction of propagation of the defect.

A photograph is taken before creating the defect and after its arrest, in order to measure the initial knit periodicity $l_x$ and $l_y$ (see Fig.~\ref{fig:fig2}(c)) and count the final number of laddered stitches $n_\mathrm{lad}$, respectively.
Four holes are pierced successively in each sample with increasing $\Delta x$ spacing, in a zigzag pattern depicted in Appendix~\ref{sec:PiercingProtocol}.
 The values of $\Delta y$ and $\Delta x$ are chosen in order to explore a wide range of configurations, such that $0.24 \leq l_x/l_0 \leq 0.45$ and $0.24 \leq l_y/l_0 \leq 0.33$ (see configuration diagram in Appendix~\ref{sec:diagram}).
In all experiments, we ensure that  $F_x$ never exceeds \SI{15}{\newton}, so that the longitudinal strain of the Nylon filament, estimated as $F_x/(n_y E\pi d^2/4)$, remains below 1$\%$, that is within the elastic regime of the material.

\begin{figure*}[ht!]
    \centering
    \includegraphics[width=0.9\linewidth]{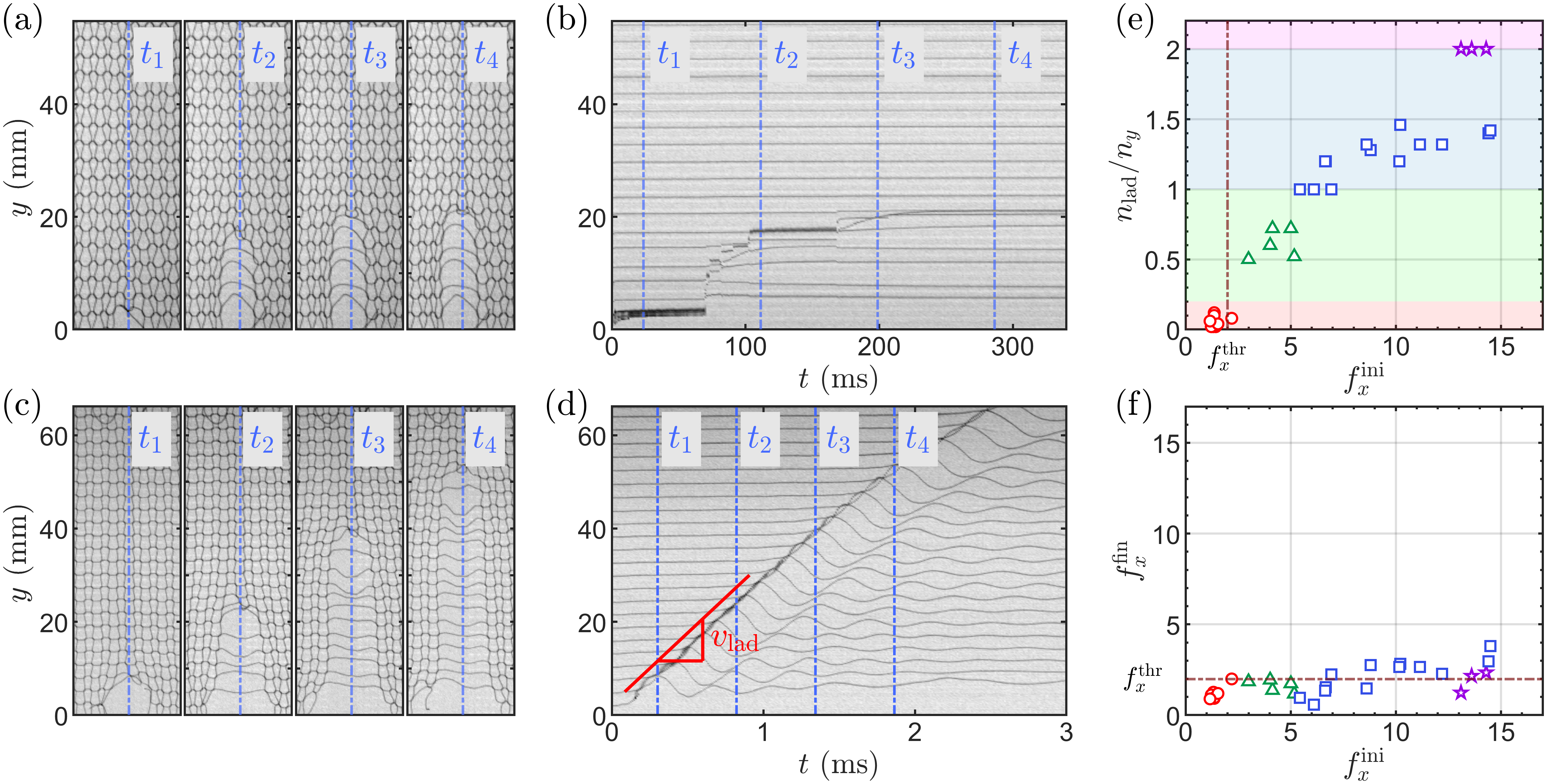}
    \caption{Comparison of laddering experiments at a low and high initial tension $f_x^\mathrm{ini} = 1.35 $ (a-b) and $f_x^\mathrm{ini} = 13.6$ (c-d), respectively. 
    (a-c) Pictures of the knit at successive instants $t_{1,...,4}$ while laddering. (b-d): Spatio-temporal diagrams of the central pixel column of the ladder (dash-dotted lines on (a) and (c)). The dash-dotted lines locate the corresponding times $t_1$ to $t_4$ on (a) and (c). The solid red line on (d) highlights the front of the defect propagation from stitch 2 to 10, its slope corresponds to the laddering velocity $v_\mathrm{lad}$. 
    %=====================================================%
    (e) Ratio of the number of laddered stitches $n_\mathrm{lad}$ to the total number of stitches in one column $n_y = 50$, as a function of $f_x^\mathrm{ini}$. Symbols indicate the final damage level: $n_\mathrm{lad}/n_y < 0.2$ (less than 10 laddered stitches, red circles), $0.2 \le n_\mathrm{lad}/n_y < 1$ (green triangles), $1 \le n_\mathrm{lad}/n_y < 2$ (blue squares), $n_\mathrm{lad}/n_y \ge 2$ (purple stars). 
    %=====================================================%
	(f) Tension in the knit just after the ladder arrest $f_x^\mathrm{fin}$ as a function of $f_x^\mathrm{ini}$. Symbols are the same as in (e). The dashed-dotted line in (e-f) indicate the threshold tension $f_x^\mathrm{thr}$.
    }
    \label{fig:fig3_results}
\end{figure*}

We also performed laddering experiments using a simulated knitted fabric. To this end, we used the model introduced in~\cite{crassous2023discrete}, which is based on Discrete Elastic Rods (DER) coupled with dry contacts. Threads are modeled as a set of point masses connected by extensional springs to account for the bending, stretching, and twisting energies of the filaments. Each mass corresponds to the center of a sphere of diameter $d$, and two successive spheres are linked with a cylinder of diameter $d$. Contact between two cylinders generates a normal force, modeled as a spring-dashpot system, and a tangential force following Coulomb's law with friction coefficient $\mu$.
The system used for the simulations is a knitted fabric composed of $42 \times 42$ jersey stitches with $d/l_0=1/70$. 
The fabric is initially placed in a particular bent configuration with $l_x/l_0=l_y/l_0=0.245$. It is then slowly stretched at $\mu=0$ (to avoid history-dependent behavior) by progressively increasing the fabric dimensions, $\Delta x$ and $\Delta y$, through imposed endpoint positions. 
Once the fabric dimensions reach some defined target values, the positions of the endpoints are kept fixed for the remainder of the simulation. The system is then allowed to relax before friction is introduced with $\mu=0.4$ and the thread is cut at the chosen location, generating the propagation of a ladder as shown in Figure~\ref{fig:fig2}(b). 

Fig.~\ref{fig:fig3_results} summarizes the results of the laddering experiments conducted for various values of $\Delta y$ and $\Delta x$. In order to facilitate the comparison between experimental and numerical results, we introduce the dimensionless force per thread $f_x = \frac{F_x}{n_y f_0}$. The force $f_0 = B \kappa_0^2=\SI{14.8}{\milli\newton}$, with $\kappa_0=2\pi/l_0$ the curvature of a circle of perimeter equal to the stitch length, is the typical bending force scale in the knitted fabric. Besides, the laddering velocity is normalized by the typical wave velocity $c_\mathrm{b} = \sqrt{\frac{f_0}{\rho_L}}=\frac{2\pi}{l_0}\sqrt{\frac{B}{\rho_L}}=\SI{30.6}{\meter\per\second}$ of a bending wave of wavelength $l_0$. 

Depending on the value of the tension applied to the knit, which acts here as a control parameter, very different behaviors are observed, ranging from no propagation, to partial and sequential propagation, or to continuous and steady propagation, as illustrated in Fig.~\ref{fig:fig3_results}(a-d). The final state at low tension (Fig.~\ref{fig:fig3_results}(a)) shows only a few destroyed cells, whereas at higher tension, the ladder reaches the top of the sample (Fig.~\ref{fig:fig3_results}(c)). The propagation dynamics are also very different. At low tension, the spatio-temporal evolution of the midline shows irregular jump-like dynamics (Fig.~\ref{fig:fig3_results}(b)), whereas propagation occurs essentially at constant velocity at high tension (Fig.~\ref{fig:fig3_results}(d)). Finally, note that the timescales of the two cases vary over almost two orders of magnitude.

In order to quantify this transition, we introduce an indicator of the final damage level, $n_\mathrm{lad}/n_y$, where $n_\mathrm{lad}$ is the total number of destroyed cells.
Therefore, $n_\mathrm{lad}/n_y$ corresponds to the column fraction untangled by the topological defect. A complete laddered column corresponds to $n_\mathrm{lad}/n_y=1$. Fig.~\ref{fig:fig3_results}(e) shows how this quantity increases with the initial tension $f_x^\mathrm{ini}$. For initial tensions below a threshold value $f_x^\mathrm{thr} \simeq 2$, there is essentially no destruction ($n_\mathrm{lad}/n_y<0.1$), and it is practically impossible to define a propagation velocity. For $f_x^\mathrm{ini}>f_x^\mathrm{thr}$, partial sub-column propagation $1>n_\mathrm{lad}/n_y>0$ or full column propagation $n_\mathrm{lad}/n_y=1$ are observed. Then, at even higher tensions, the second downward column is entirely laddered, and eventually a second column of stitch ($n_\mathrm{lad}/n_y=2$) is laddered.

Another quantity of interest is the final tension $f_x^\mathrm{fin}$ after the laddering stopped, displayed in Fig.~\ref{fig:fig3_results}(f). The final force $f_x^\mathrm{fin}$ is constant regardless of the destruction level, and is also equal to $f_x^\mathrm{thr}$. The threshold force $f_x^\mathrm{thr}$ is then the value to which the force relaxes once defect propagation is complete: the knit continues to ladder as long as $f_x - f_x^\mathrm{thr} > 0$. The threshold force $f_x^\mathrm{thr}$ can be interpreted as a pinning force which retain the free loop, noted (1) in Fig.~\ref{fig:fig2}(d), which is submitted to the driving force $f_x$. We take for $f_x^\mathrm{thr}$ the mean value of $f_x^\mathrm{fin}$ for all experiments: $f_x^\mathrm{thr}=2.0\pm 0.8$. This value may be linked to the natural curvature $\bar{\kappa}$ of the thread, due to the knitting step, in which the applied stresses exceed the plasticity threshold of the Nylon thread. Taking $B \bar{\kappa}^2$ for the pinning force, we have $f_x^\mathrm{thr} = (\bar{\kappa}/\kappa_0)^2$. From the measured values of $\bar{\kappa}$ (see Appendix~\ref{sec:curvature}), the threshold $f_x^\mathrm{thr}=2$ falls between the average and the maximal values of $(\bar{\kappa}/\kappa_0)^2$. We have checked that in numerical simulations, where the natural curvature of thread is $\bar{\kappa}=0$, the force threshold is zero as expected (see simulation snapshots in Appendix~\ref{sec:simuls}).

\begin{figure}[t]
    \centering
    \includegraphics[width=0.9\linewidth]{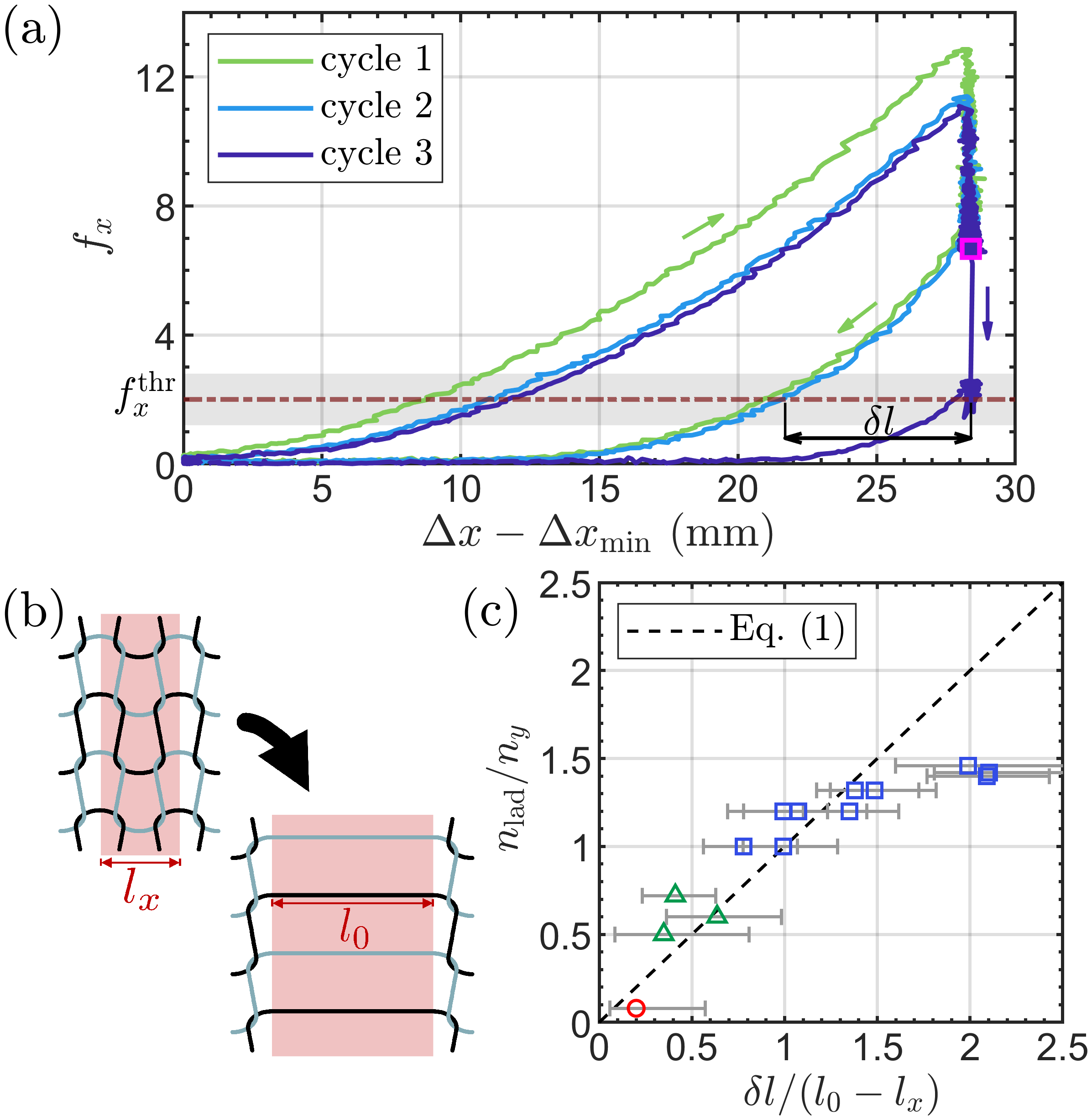}
    \caption{(a) Force-displacement profile of the knitted fabric. Grey shaded area indicates the standard deviation of the laddering threshold $f_x^\mathrm{thr}$ (dashed-dotted line). Magenta square: $f_x^\mathrm{ini}$ before cutting the thread in cycle 3. (b) Sketch comparing the initial width of the stitch with the width of the ladder. The released length is $l_0 - l_x$. (c) Comparison of the measured column ratio of laddered stitches with the prediction of equation~\eqref {eq:rel_nDS_deltal}. Errors bars are evaluated from the standard deviation of $f_x^\mathrm{thr}$.}
    \label{fig:fig4}
\end{figure}

Since the threshold force $f_x^\mathrm{thr}$ governs the relaxation of the system, one may expect the destruction level to be directly related to $f_x^\mathrm{thr}$. To make this relation explicit, we performed experiments to observe the retraction of an undamaged knit from a stretched configuration at $f_x^\mathrm{ini} > f_x^\mathrm{thr}$ to a force $f_x=f_x^\mathrm{thr}$.
The experiment is schematically described in Fig.~\ref{fig:fig4}(a) and in Appendix~\ref{sec:StretchingProctocols}. Three deformation cycles up to a maximal stretch $\Delta x_\mathrm{max}$ are applied to an undamaged knit, and the force-displacement profiles collapse onto a single cycle from the first retraction. During the third cycle, a stitch is cut at $\Delta x_\mathrm{max}$: as expected the force relaxes abruptly to a value $\simeq f_x^\mathrm{thr}$. The distance  denoted $\delta l$ in Fig.~\ref{fig:fig4}(a) represents the shortening that an {\it undamaged} knit would need to undergo to bring its force from $f_x^\mathrm{ini}$ back to $f_x^\mathrm{thr}$. We can simply express this displacement $\delta l$ in terms of the average number of columns that must be destroyed: destroying one column releases a length $(l_0-l_x)$ (see Fig.~\ref{fig:fig4}(b)), and therefore the number of destroyed columns is :
\begin{equation}\label{eq:rel_nDS_deltal}
    \frac{n_\mathrm{lad}}{n_y} = \frac{\delta l}{l_0 - l_x}.
\end{equation}

Eq.~\eqref{eq:rel_nDS_deltal} is tested in Fig.~\ref{fig:fig4}(c). For this, $\delta l$ is measured using the protocol described in Fig.~\ref{fig:fig4}(a) for various values of $\Delta y$ and $\Delta x_\mathrm{max}$, while $l_x$ is obtained from a picture taken at $\Delta x_\mathrm{max}$ before cutting the thread. The predicted damage $\delta l/(l_0 - l_x)$ is then compared with the real amount of damage $n_\mathrm{lad}/n_y$ observed after cutting the thread at $\Delta x_\mathrm{max}$ during the third cycle. 
The agreement is good up to the value 1.5, beyond which the cut thread prevents the start of a second upward-propagating ladder by remaining stuck in loops (4) and (5) depicted in Fig.~\ref{fig:fig2}(d) (see Appendix~\ref{sec:pinning}).

\begin{figure}[ht]
    %\centering
    \includegraphics[width=0.9\linewidth]{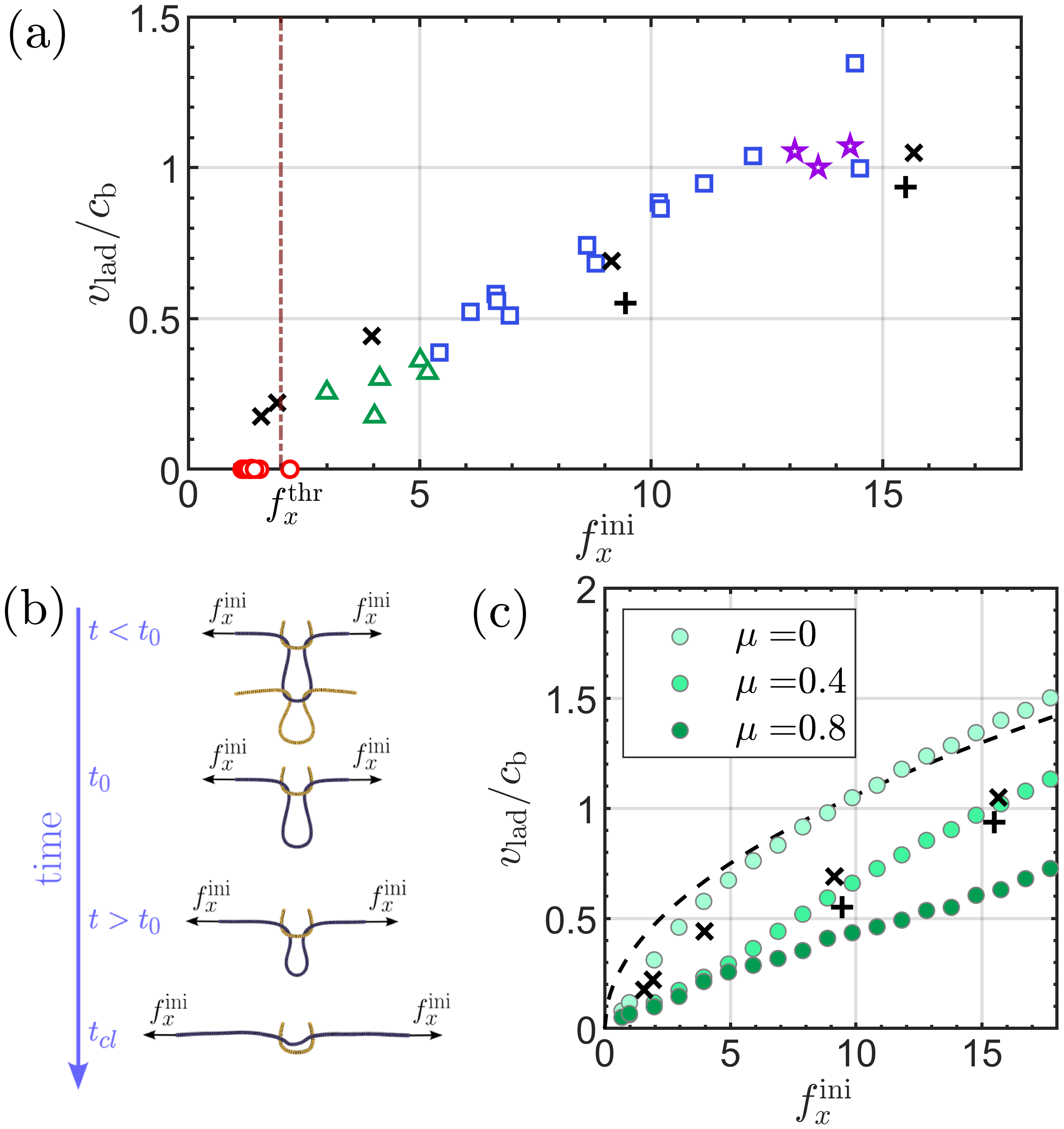}
    \caption{
    (a) Normalized laddering velocity as a function of the initial tension $f_x^\mathrm{ini}$. Experimental data are plotted with the same symbols as in Fig.~\ref{fig:fig3_results}(e). Black crosses are DER results for a $42\times42$ knit with $\mu=0.4$, at several values of $l_x/l_0$ with $l_y/l_0 = 0.25$ ($\times$) and $l_y/l_0 = 0.302$ ($+$).
    (b) Snapshots of the single stitch simulation at various instants. The laddering time is $\tau = t_{cl} - t_0$.
    (c) Normalized laddering velocity as a function of the applied tension for different friction coefficient $\mu$ in the single stitch simulation (circles). Black crosses are the DER results for a $42\times42$ knit as in (a). The dashed line displays $\beta\sqrt{f_x^\mathrm{ini}}$ with $\beta = 0.34$.}
    \label{fig:fig5}
\end{figure}

We now discuss the laddering limit in knits under applied forces $f_x^\mathrm{ini}$ such that a well-defined continuous unraveling velocity exists (see Fig.~\ref{fig:fig3_results}(d)). Fig.~\ref{fig:fig5}(a) shows the variation of the laddering velocity $v_\mathrm{lad}$ as a function of tension. The velocities obtained for knits with very different aspect ratios depend only on the initial tension $f_x^\mathrm{ini}$. As soon as the tension exceeds the threshold $f_x^\mathrm{thr}$, the velocity varies linearly with the force. We also note an excellent agreement between experimental data and numerical simulations using frictional elastic rods. The linearity suggests that the unraveling dynamics are not governed by inertia. Indeed, in that case one would expect the velocity $v_\mathrm{lad}$ to depend only on the tension $f_x^\mathrm{ini}$ (at least in the regime $f_x^\mathrm{ini} >> f_x^\mathrm{thr}$) and on the linear mass density $\rho_L$, leading to $v_\mathrm{lad} \sim \sqrt{f_x^\mathrm{ini} / \rho_L}$. This is clearly not the observed behavior, neither qualitatively ($v_\mathrm{lad}\sim f_x^\mathrm{ini}$) nor quantitatively. On the contrary, the linearity suggests the presence of a dissipating force in the system. The obvious source of dissipation lies in frictional forces at contact points between the threads. To demonstrate the importance of dissipating forces, we simplified the problem to the study of the dynamics of a single numerical loop. We consider an elastic loop initially hooked to a lower loop (see Figure~\ref{fig:fig5}(b)). The lower loop is removed, and we measure the time $\tau$ required for the upper contact to be released (defined as the instant when the contact between loops of rows $n$ and $n+1$ breaks). The propagation velocity is then $v_\mathrm{lad}=l_y/\tau$. Fig.~\ref{fig:fig5}(c) shows the evolution of $v_\mathrm{lad}$ for this simplified system as a function of the tension $f_x$ for a frictionless system ($\mu = 0$) and for two frictional systems $\mu = 0.4$ and $\mu = 0.8$. First, the velocity measurements for the full knits agree very well with this simple single-loop model for $\mu = 0.4$ without adjustable parameters. Second, it is clear that the lower the friction, the higher the unraveling velocity. Finally, the presence of friction linearizes the dependence $v_\mathrm{lad}(f_x^\mathrm{ini})$. In the absence of friction, the dynamics are inertial, with $v_\mathrm{lad}\sim \sqrt{f_x^\mathrm{ini}}$ (see dashed line in Fig.~\ref{fig:fig5}(c)). 
The observed linear behavior $v_\mathrm{lad}\sim f_x^\mathrm{ini}$ is therefore due to the presence of a characteristic force $f_e$ related to elasticity and friction. In the case of small extensions, one expects $f_e = \mu B \kappa_c^2/f_0 = \mu  \kappa_c^2/\kappa_0^2 $, with $\kappa_c$ a characteristic curvature such that $\kappa_0 = 2\pi/l_0 < \kappa_c < 1/d$. Writing the velocity as $v_\mathrm{lad}(f_x^\mathrm{ini}, f_e, \rho_L)$ and imposing $v_\mathrm{lad}\propto f_x^\mathrm{ini}$ as observed experimentally, one obtains $v_\mathrm{lad}(f_x^\mathrm{ini}, f_e, \rho_L) \sim \sqrt{(f_x^\mathrm{ini})^2 / \rho_L f_e}$, up to a prefactor of order unity. This relation can be rewritten as
\begin{equation}
    \frac{v_\mathrm{lad}}{c_\mathrm{b}} \sim \frac{\kappa_0}{\sqrt{\mu} \kappa_c} f_x^\mathrm{ini}
\end{equation}
with $0.14 < \kappa_0 / \sqrt{\mu} \kappa_c  < 1.6 $. This relation is in good agreement with the experimental data, $v_\mathrm{lad} / c_b^0 \simeq 0.1 f_x^\mathrm{ini}$. In other words, the hypothesis that bending elasticity combined with friction controls the laddering velocity is consistent with all our results. A more in-depth study of this complex dynamics and the interplay between tension, curvature, and friction in the deployment of the stitch will be the subject of future work.

Summarising, we have uncovered the control parameter of the laddering process in a pre-stretched fabric. A single threshold tension $f_x^{\mathrm{thr}}$ controls both the onset and the end of the propagation of such topological defect. $f_x^{\mathrm{thr}}$ is not due to fiber-fiber friction but arises from the natural curvature of the thread, which is induced by the knitting process and, in practice, can be reinforced by physico-chemical blocking processing~\cite{Choi2003}.
During the defect propagation from stitch to stitch, the tension of the fabric drops due to the excess  length of thread liberated by the untangling of the loops. At moderate tension, the final size of the ladder $n_\mathrm{lad}$ can be predicted from the value of the tension threshold $f_x^{\mathrm{thr}}$, and the mechanical and geometrical knowledge on the pristine knit. However at high tension, it is limited by the dynamics of the cut thread, that frees only a limited number of loops in the course direction. Furthermore, we evidenced that the velocity of bending waves of wavelength $l_0$ sets the order of magnitude of the laddering velocity $v_\mathrm{lad}$. Using simplified single-loop simulations, the friction between the threads was shown to play a crucial role both on the value of the laddering velocity and on its linear scaling behavior with the initial tension $f_x^{\mathrm{ini}}$. It would be interesting to investigate in the future how laddering is affected by the transition from a loose tight-like structure to a jammed sweater-like one~\cite{Matsumoto2025} by varying the ratio $d/l_0$.

Finally, we point out that although the phenomenon of laddering resembles crack propagation, unlike cracking that usually splits a material in two separate parts, laddering preserves the integrity of the structure which remains held by the ladder bars. Therefore, laddering can be viewed as a damage-control and damage-mitigation mechanism that can be used positively for energy absorption or deployment of structures. Owing to the peculiar structure of weft knitted fabrics, the direction of propagation of a topology defect is perfectly controlled by the pattern of the fabric~\cite{Shimamoto2025}. In Jersey, a ladder defect will propagate along the wale direction freeing extra thread length in the course direction. Moreover, following a local break of the thread, the fabric stress will drop sharply thanks to the rapid liberation of this extra thread length, thus efficiently preventing further thread breakage. Therefore, the control of direction and propagation threshold of ladder defects might be harnessed to conceive protective knitted structures, bearing a network of pre-designed topological weaknesses, in order to protect a frail object against stresses exceeding the ladder propagation threshold. Finally, a ladder defect might not look nice, but it actually helps the fabric hold together against brutal events. The laddering mechanism could thus inspire damage-mitigation strategies in other types of architected materials, for example by designing length reservoirs in their structure~\cite{Zou2021}.

%%%%%%%%%%%%%%%%%%%%%%%%%%%%%%%%%%%%%%%%
\begin{acknowledgments}
A.S. acknowledges L. Babin, P.A. Langrognet, R. Lautier and S. Cornab\'e for preliminary work. A.S and A.F. thank D. Le Tourneau, P. Metz and V. Coullon for their help in building the experimental setup, and V. Vidal for her curvature computation code. The authors thank J.C. G\'eminard for feedback on the manuscript. 
This work was supported by Agence Nationale de la Recherche Grant ANR-23-CE30-0015.
\end{acknowledgments}
%%%%%%%%%%%%%%%%%%%%%%%%%%%%%%%%%%%%%%%%

\bibliography{refs}

\newpage

\appendix

%\begin{document}

%\renewcommand\thesection{S\arabic{section}}
%\setcounter{section}{0}

%\renewcommand\theequation{S\arabic{equation}}
%\setcounter{equation}{0}

%\renewcommand\thefigure{S\arabic{figure}}
%\setcounter{figure}{0}

%\renewcommand\thetable{S\arabic{table}}
%\setcounter{table}{0}

\section{Description of Supplementary Movies}\label{sec:movies}

We provide two movies of propagating ladder defects: in commercial tights and the experimental Jersey knitted fabric.

\begin{itemize}
    \item \textbf{Movie SM1\_DIM\_tights.mp4:} Laddering experiment performed on commercial tights (Mes Essentiels, DIM) captured at 110000 frames per second, here shown at 15 frames per second. The initial hole was made using a thread burner.
    
    \item \textbf{Movie SM2\_exp\_knit.mp4:} Laddering experiment performed on the model Jersey knitted fabric captured at 96000 frames per second, here shown at 30 frames per second. The movie corresponds to the experiment illustrated in Fig.~\ref{fig:fig3_results}(c,d) with $f_x^\mathrm{ini} = 13.6$. The initial hole was made with a soldering iron. 

\end{itemize}

\section{Details on experimental protocols}\label{sec:ExpProtocols}

\subsection{Stretching protocols}\label{sec:StretchingProctocols}

\par We now detail the stretching steps prior to making a hole. 
The first step is to bring the knitted fabric from the loose configuration in which it is mounted on the frame of the bi-axial tensile machine, to the starting configuration at the target $\Delta y$ spacing and a $\Delta x_\mathrm{min}$ reference spacing. 
From the initial loose configuration, a stretching motion is imposed in the $y$ direction up to the target $\Delta y$ while keeping the $\Delta x$ distance low enough to maintain the forces $F_x$ and $F_y$ close to zero during this stretching step. 
The $\Delta y$ spacing is then kept constant, while a force-controlled stretching motion is applied in the $x$ direction toward the target minimal tension of $F_{x\mathrm{, min}} = \SI{0.5}{\newton}$ that sets the initial configuration. This minimal tension is necessary by construction, to keep the holding paperclips slightly under tension, compensating the effect of gravity.
When the knit has reached the target tension, a gentle stroke is performed on the paperclips and on the knit itself with a rounded tool in order to relax the frictional contact points, while the feedback loop on the force continues slightly stretching the knit in the $x$ direction to maintain the tension at $F_{x \mathrm{, min}}$. Finally, the $x$ stretching motion is stopped, and the stretching position defines the reference spacing $\Delta x_\mathrm{min}$.

\begin{figure*}[ht]
    \centering
    \includegraphics[width=0.99\linewidth]{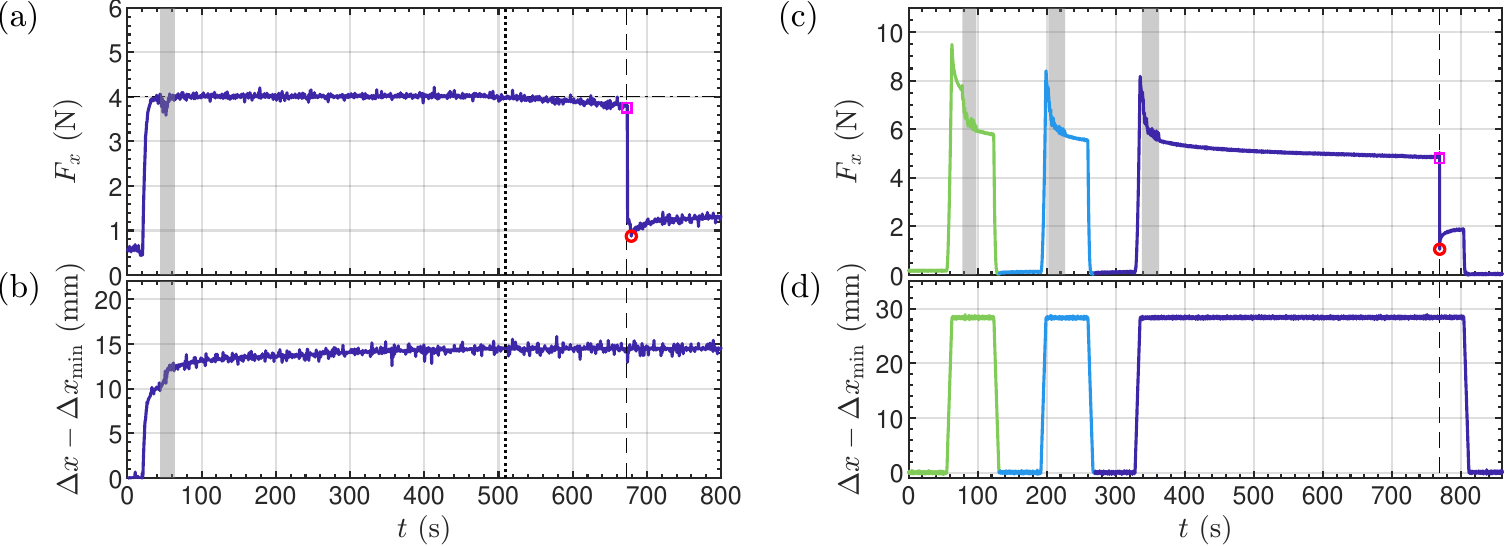}
    \caption{(a)(c) Forces $F_x$ and (b)(d) displacements $\Delta x - \Delta x_\mathrm{min}$, as functions of time $t$ during a laddering experiment performed following protocol A (a-b) and B (c-d). The gray-shaded regions show the time intervals during which the knit is gently stroked to relax the contacts. In panels (a-b), the horizontal dashed-dotted line locates the target force $F_{x \mathrm{, max}}$, and the vertical dotted line the time at which the frame is fixed in place. In all panels, the vertical dashed line corresponds to the time at which a defect is created in the knit.
	The laddering force $F_x^\mathrm{ini}$ is indicated by a magenta square and the stopping force $F_x^\mathrm{fin}$ by a red circle.}
    \label{fig:stretching_protocols}
\end{figure*}

The second step is to stretch the fabric in the $x$ direction up to the spacing $\Delta x_\mathrm{max}$ at which a hole is pierced. Two different protocols, named A and B, were used for this second step. Fig.~\ref{fig:stretching_protocols} depicts the force $F_x$ and displacement $\Delta x - \Delta x_\mathrm{min}$ as function of time $t $ for both protocols, where $t=0$ is between the end of the first step and the beginning of the second step.

\begin{itemize}
    \item  \textbf{Protocol A:} the knit is stretched by a second force-controlled motion in the $x$ direction, toward a target force $F_{x \mathrm{, max}}$ indicated by a dashed-dotted horizontal line on Fig.~\ref{fig:stretching_protocols}(a).
Once again, when the force $F_{x \mathrm{, max}}$ is reached, the knit and paperclips are gently stroked while the feedback loop maintains the tension $F_{x \mathrm{, max}}$ by slightly increasing $\Delta x$. The stroking step is indicated by the grey area on Fig.~\ref{fig:stretching_protocols}(a) and (b).
The feedback loop remains active during the time the soldering iron is placed in front of the desired hole position using a micromanipulator and heated up. Afterwards, the motion of the bars in $x$ direction is stopped (see vertical dotted line on Fig.~\ref{fig:stretching_protocols}(a) and (b)). 

\item \textbf{Protocol B:} was designed to test the damage prediction of equation~\ref{eq:rel_nDS_deltal}, by measuring $\delta l$ as shown in Fig.~\ref{fig:fig4}(a). Instead of a target force $F_{x \mathrm{, max}}$ like in protocol A, we impose a target spacing $\Delta x_\mathrm{max}$.
As shown on Fig.~\ref{fig:stretching_protocols}(c-d), three complete loading cycles are performed between $\Delta x_\mathrm{min}$ and $\Delta x_\mathrm{max}$ at constant velocity $v = \SI{2}{\milli\meter\per\second}$. In each cycle, a waiting time is imposed at $\Delta x_\mathrm{max}$, during which the knitted sample and the paper clips are gently stroked (grey areas in Fig.~\ref{fig:stretching_protocols}(c-d)) in order to speed up the relaxation dynamics of the fabric and reach a stabler force level. Actually, the waiting time and stroking at $\Delta x_\mathrm{max}$ during the first two cycles are done in order to mimic the last cycle which is the laddering cycle, where the waiting time is necessary to install the soldering iron to pierce the hole. 
\end{itemize}

After stopping the motion of the bars in protocol A and positioning the soldering iron in protocol B, a photograph of the initial rest state of the knit is taken in order to measure the initial spatial periodicity of the knit $l_x$ and $l_y$, a high-speed video recording is started to measure the laddering velocity $v_\mathrm{lad}$ and the thread is burned with the tip of a soldering iron at the planned location. The creation of a topological defect triggers a sudden drop in the force $F_x$, highlighted by a vertical dashed line on Fig.~\ref{fig:stretching_protocols}(a-d). The initial and final forces $F_x^\mathrm{ini}$ and $F_x^\mathrm{fin}$ are defined just before and just after this sharp force drop, as depicted respectively by the magenta square and the red circle on Fig.~\ref{fig:stretching_protocols} (a) and (c). Note that the force at which the knit ladders $F_x^\mathrm{ini}$ is slightly below the target force $F_{x \mathrm{, max}}$ in protocol A because of stress relaxation of the fabric. A final photograph is taken after laddering stopped to measure the final number of destroyed cells $n_\mathrm{lad}$, before bringing the $\Delta x$ spacing back to $\Delta x_\mathrm{min}$. 

To finish, the topology defect is stabilised against further propagation by holding the cut threads using small lightweight Nylon nuts and screws. Ladder tips that haven't reached the sample border are stabilised by holding the free end loops with small safety pins.

\subsection{Piercing protocol}\label{sec:PiercingProtocol}
\begin{figure}[ht]
    \centering
    \includegraphics[width=0.98\linewidth]{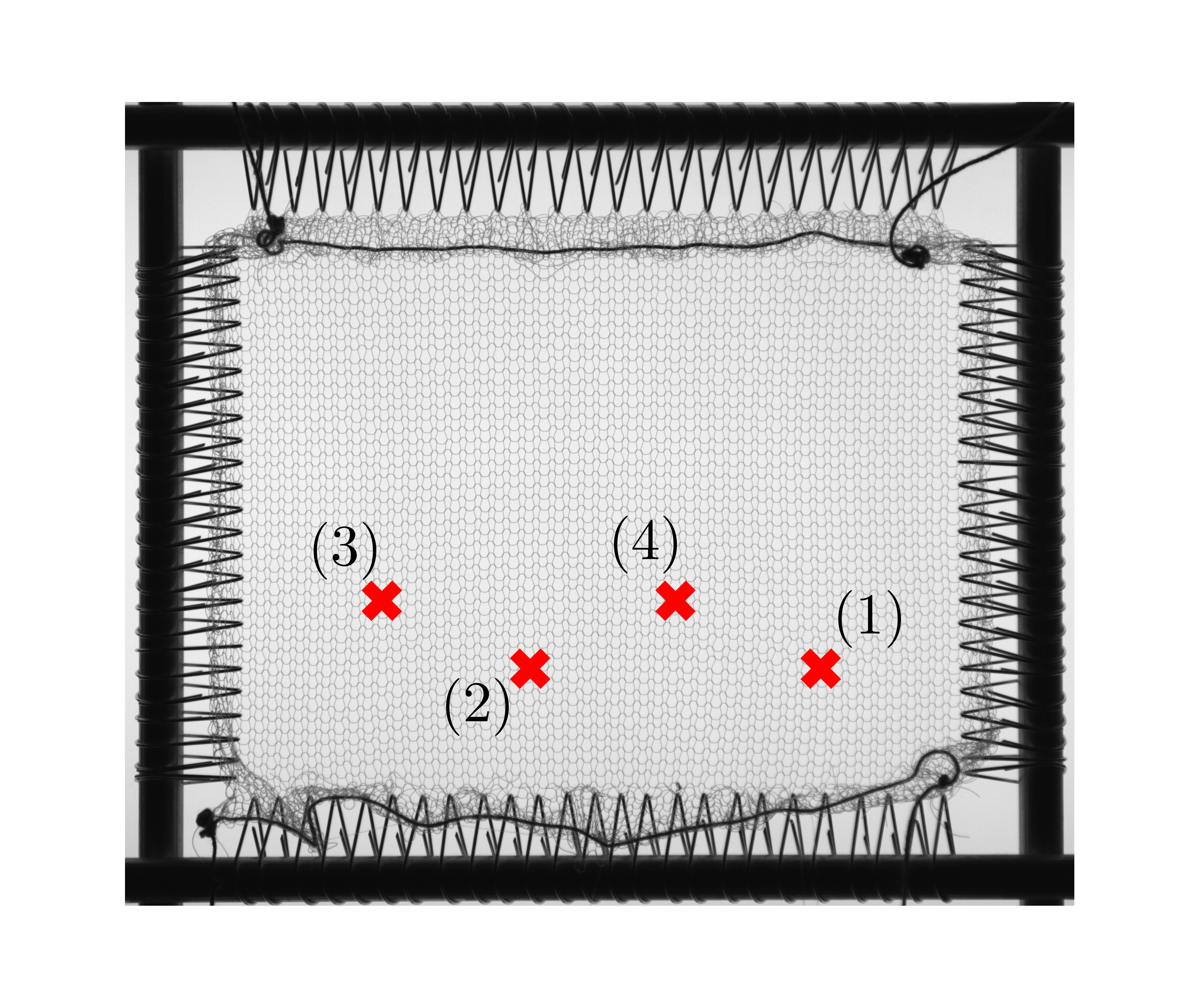}
    \caption{Photograph of the knitted sample indicating the location of the holes. The numbers indicate the order in which they are pierced.}
    \label{fig:zig_zag_hole}
\end{figure}

On most knitted samples, four laddering experiments are performed at the same vertical stretching $\Delta y$ and with increasing initial force $F_x^\mathrm{ini}$. The topology detects are created by burning rather than cutting the thread in order to introduce minimal additional stress. The four holes are made in a zigzag pattern as shown on Fig.~\ref{fig:zig_zag_hole}, in order to minimize the interaction between the defects. The space between each hole is ten stitches horizontally and five vertically.
This zigzag pattern means that the column fraction of laddered stitches $n_\mathrm{lad}/n_y$ is slightly different between the lower holes (1-2) and the upper holes (3-4) for a complete upward ladder starting from the broken stitch with no downward laddering stitches.

One sample was used specifically to complete the line of non-propagating topology defects for $F_x^\mathrm{ini}$ below the laddering threshold in the configuration diagram (red circles in Fig.~\ref{fig:diagram}). In this sample, seven holes were pierced at increasing vertical stretching $\Delta y$ with constant $F_{x \mathrm{, max}} =\SI{1}{\newton}$, corresponding to $f_{x\mathrm{, max}}=\frac{F_{x_\mathrm{, max}}}{n_y f_0}=1.35$ per row in dimensionless units.

\section{Configuration diagram}\label{sec:diagram}
\par
\begin{figure}[ht]
    \centering
    \includegraphics[width=0.95\linewidth]{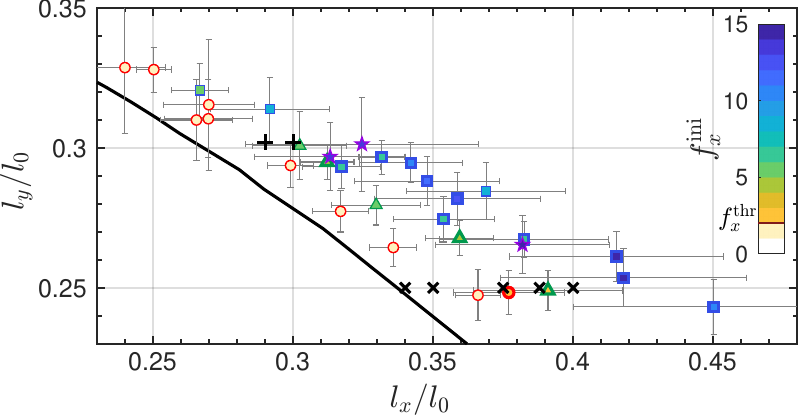}
    \caption{Configuration diagram: knit periodicity in $y$ direction $l_y/l_0$ versus the one in $x$ direction $l_x/l_0$ just before creating a topology defect. 
    Each symbol corresponds to one laddering experiment. 
    Marker face colors depicts the initial tension $f_x^\mathrm{ini}$ for experimental data, according to the color bar. Thin (resp. thick) symbol edges: data taken following protocol A (resp. B).
    Symbols shapes and edge colors are the same as in Fig.~\ref{fig:fig3_results}(e); they indicate the amount of final damage. DER simulations on $42 \times 42$ stitches are shown by black symbols: $l_y/l_0 = 0.302$ ($+$), $l_y/l_0 = 0.25$ ($\times$). The black line locates the $f_x =0$ curve computed using DER simulation on a single periodized stitch as in \cite{Crassous2024}, with $d/l_0= 1/70$.}
    \label{fig:diagram}
\end{figure}

 Fig.~\ref{fig:diagram} summarizes the initial state of the knitted fabric, in the configuration diagram representing the knit periodicity $l_x$ and $l_y$ in the $x$ and $y$ directions respectively, normalized by the stitch length $l_0$. 

 The position of each experimental data points in the configuration diagram locates the mean value of  in the undamaged zones of the knitted sample, measured using 2D Fast Fourier Transform (2D-FFT) on the photograph taken just before a creating topological defect. The error bars represent the dispersion of stitch shapes within the fabric, obtained from the width of the 2D-FFT peaks. It increases with the total damage induced by successive laddering experiments in the same sample. Additionally, the dimensionless initial tension per row $f_x^\mathrm{ini}=\frac{F_x^\mathrm{ini}}{n_y f_0}$ of the sample is color-coded on the symbol's face. The thickness of its edge line differentiates data acquired using protocol A (thin lines) and B (thick lines). 
 The target  $l_x/l_0$ and $l_y/l_0$ values of DER simulations of $42 \times 42$ Jersey stitches are located by black crosses. 
 
 The black line locates the $f_x =0$ curve computed using DER simulation on a single periodized stitch as in \cite{Crassous2024}, with $d/l_0= 1/70$. The final damage level of the experimental sample, indicated by the markers shape and edge color according to the same code as in Fig.~\ref{fig:fig3_results}(e), increases with the distance to the $f_x =0$ curve.
 
\section{Natural curvature of the thread in the experimental knit}\label{sec:curvature}

When a thread is pulled out of a knitted fabric, everyday life experience shows that it keeps a memory of the shape of the stitches, through a non-zero natural curvature. A natural curvature is induced by the knitting process, in which the imposed strain goes beyond the elasticity limit of the thread. In real fabrics, it is further modified and reinforced by a physico-chemical blocking process at the desired rest shape of the stitches.

\begin{figure}[htbp]
    \centering
    \includegraphics[width=0.98\linewidth]{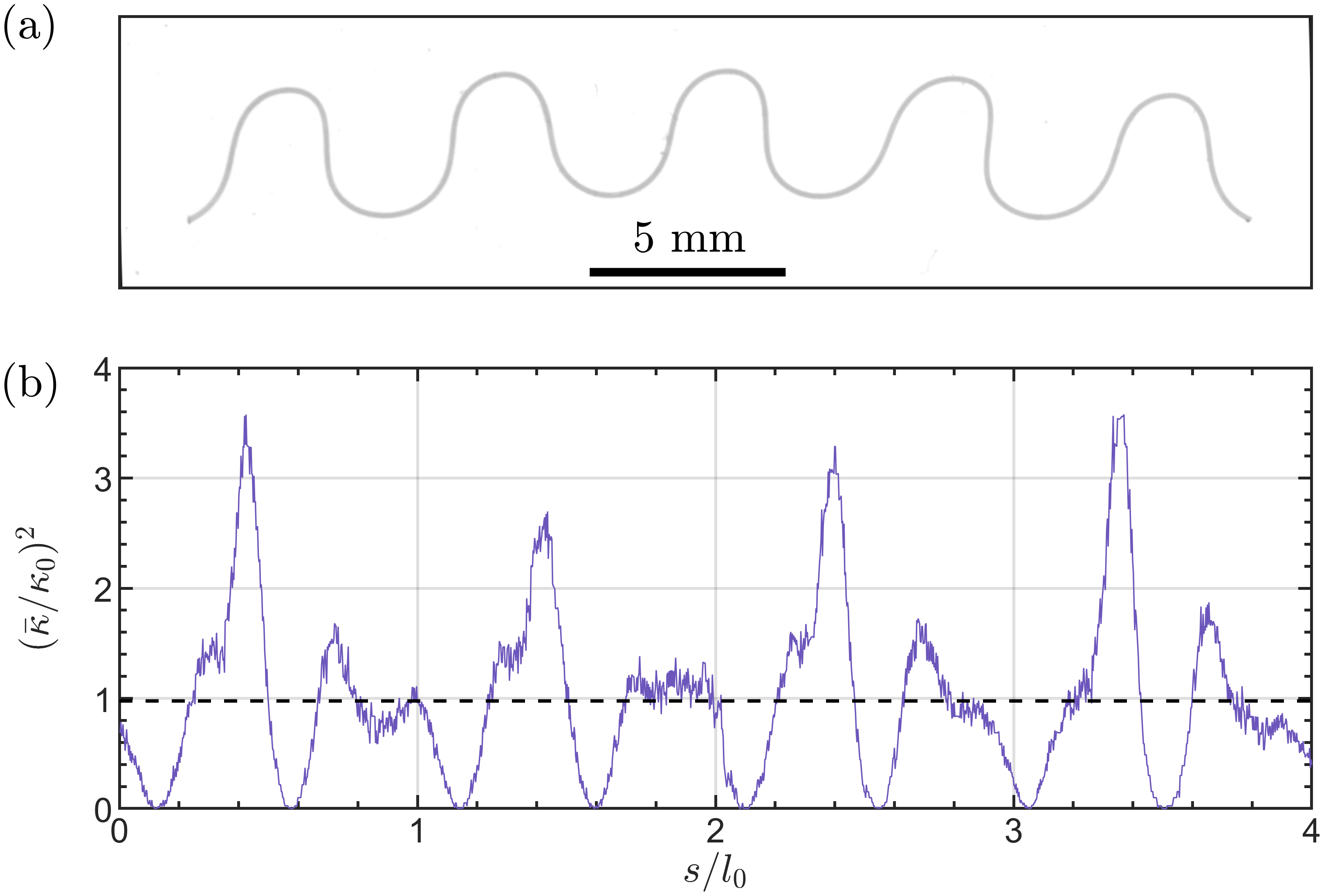}
    \caption{Residual curvature of a thread extracted from an experimental knit. (a) Photograph of the extracted thread. (b) Square of the normalized residual curvature $\bar{\kappa}/\kappa_0$ as a function of the normalized curvilinear abscissa $s/l_0$. The dashed line indicates the mean value of $(\bar{\kappa}/\kappa_0)^2$ computed over the length of four stitches.}
    \label{fig:curvature}
\end{figure}

In our model Jersey fabric, the natural curvature $\bar{\kappa}$ comes from the knitting process only. To characterize it, a thread was carefully extracted from a knitted sample by cutting away the surrounding loops; a photograph of that thread is shown in Fig.~\ref{fig:curvature}(a). The local curvature $\bar{\kappa}$ of the thread's skeleton was then measured as a function of its curvilinear abscissa $s$. 

In our dimensionless units, the force scale necessary to straighten the thread is $(\bar{\kappa}/\kappa_0)^2$, where $\kappa_0 = 2\pi/l_0$. Fig.~\ref{fig:curvature}(b) shows this quantity as a function of the normalized curvilinear abscissa $s/l_0$ over the length of four stitches. The dimensionless experimental laddering threshold $f_x^\mathrm{thr}=2$ is between the average value of $(\bar{\kappa}/\kappa_0)^2 \simeq 1$ (dashed line on Fig.~\ref{fig:curvature}(b) and is maximal value around 3.5. 

\section{Hindered laddering due to pinning of the cut thread}\label{sec:pinning}
In this section, we discuss the limits of the damage prediction by equation~\eqref {eq:rel_nDS_deltal}. In Fig.~\ref{fig:fig4}(c), the data points at high initial tension which were predicted to ladder two complete columns ($\delta l/(l_0 -l_x) \simeq 2$ actually laddered only one complete upward ladder starting from loop (1) and two complete downward ladder starting from loops (2) and (3) (see Fig.~\ref{fig:fig2}(d) for loop numbering), leading to $n_\mathrm{lad}/n_y \simeq 1.5$. 

\begin{figure}[htbp]
    \centering
    \includegraphics[width=0.95\linewidth]{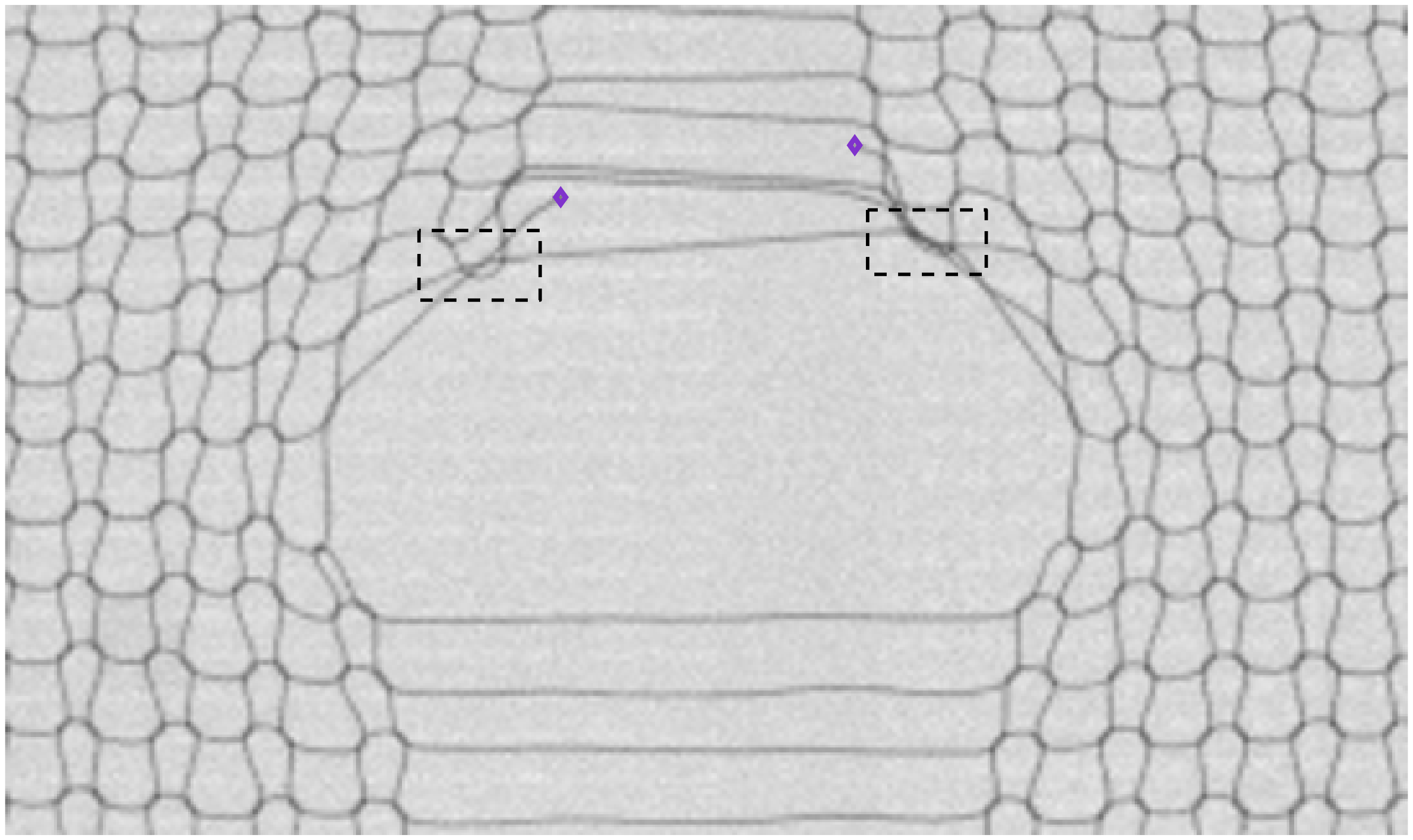}
    \caption{Photograph of the final state of the fabric after laddering for a sample with initial tension $f_x^\mathrm{ini} = 14.51$ and $\delta l/(l_0 -l_x) \simeq 2$. This picture, focused on the hole region, shows hindered laddering (one upward ladder and two downward ladders instead of two upward and two downward ladders), due to pinching of the cut thread in the side loops located by dashed boxes. Purple diamonds symbols highlight the two ends of the cut thread.}
    \label{fig:S5_pinched_thread}
\end{figure}

This discrepancy can be explained by observing the final state of the knit in the example shown on Fig.~\ref{fig:S5_pinched_thread}, with the initial tension $f_x^\mathrm{ini} = 14.51$. On this photograph, we can see that the second upward ladder that was expected to propagate was hindered because the cut thread got pinched in loops (4) and (5) (dashed rectangles) so that neither of them was free to unwrap and initiate a new ladder. As a result, the final tension remained well above the laddering threshold : $f_x^\mathrm{fin} = 3.81$ in this experiment.

Predicting the maximal number of laddered columns at high initial tension requires to model both the retraction of the cut thread through several interlacing regions and the one of the loops around the initial hole.

\section{Simulations}\label{sec:simuls}

\begin{figure}[ht]
    \centering
    \includegraphics[width=\linewidth]{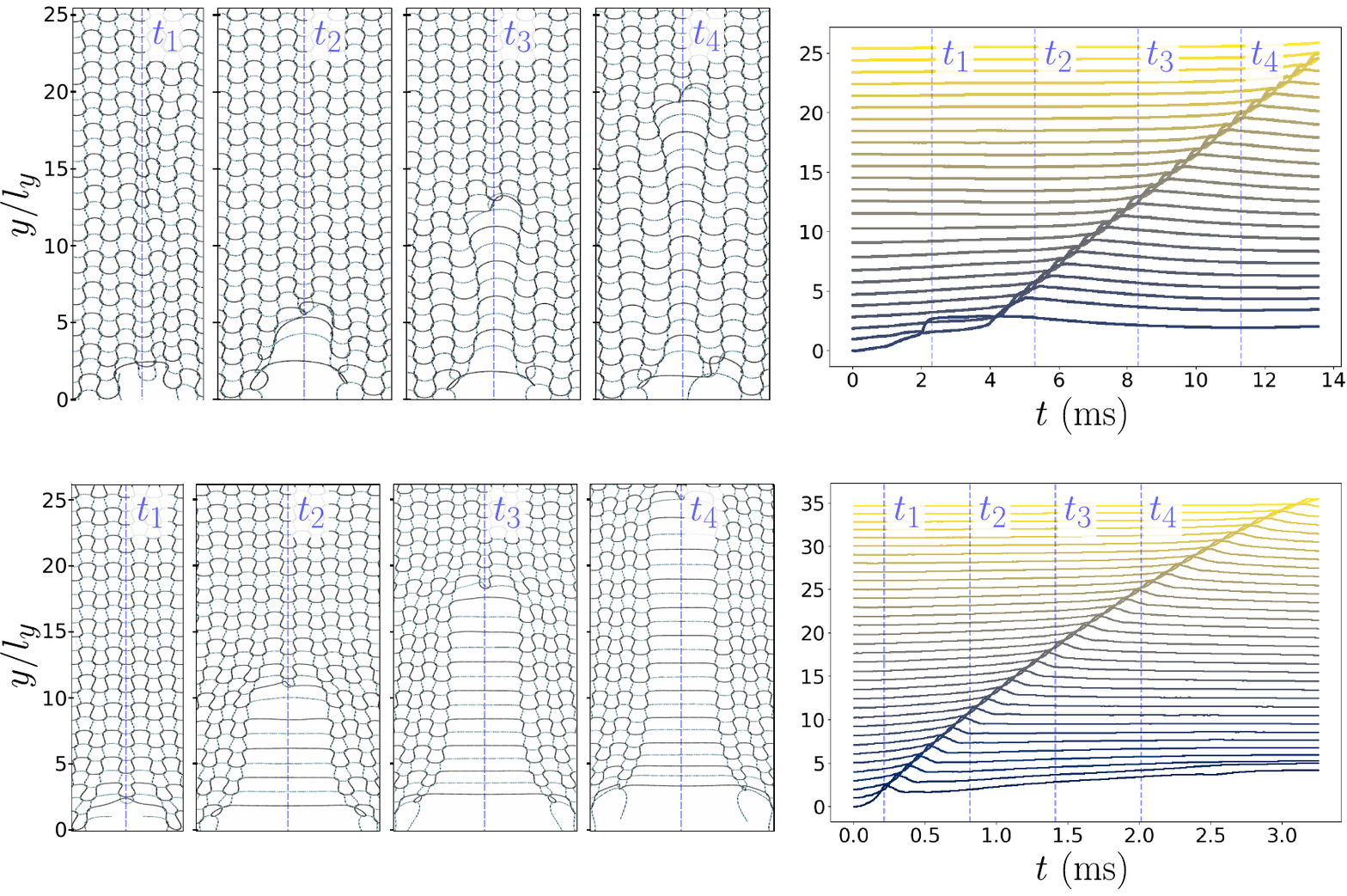}
    \caption{Comparison of laddering DER simulations performed at a low initial tension $f_x^\mathrm{ini}=1.58$ (top) and a high initial tension $f_x^\mathrm{ini}=15.67$ (bottom). Left: snapshots of the simulated knits at successive instants $t_{1,...,4}$. Right: spatio-temporal diagrams of the central point column of the ladder.}
    \label{fig:spatio_sim}
\end{figure}

Fig.~\ref{fig:spatio_sim} shows snapshots of two simulated laddering fabrics at four evenly spaced instants $t_{1,...,4}$, along with the time evolution of the $y$ position of the middle point of the laddering column. In the top panel, the simulation was conducted with a low initial tension, $f_x^\mathrm{ini}=1.58$, which is below the average threshold value $f_x^\mathrm{thr}=2.0$ below which the experimental knit does not ladder.  
In the bottom panel, the simulation was conducted at high initial tension, $f_x^\mathrm{ini}=15.67$. 

We see that the simulated knitted fabric still ladders at $f_x^\mathrm{ini}<f_x^\mathrm{thr}$. This is expected if the laddering threshold is due to a non zero natural curvature in the experimental knit (see Section~\ref{sec:curvature}), whereas the natural curvature is equal to zero in the DER simulations. Still, the laddering velocity is roughly ten times slower than in the high-tension case ($v_ \mathrm{lad}/c_b^0=0.176$ for $f_x^\mathrm{ini}=1.58$, against $v_ \mathrm{lad}/c_b^0=1.05$ for $f_x^\mathrm{ini}=15.67$). This reduced speed can be understood intuitively from the snapshots in the top panel. To free itself from the upper loop, each loop has less space than in the high-tension case and must therefore twist and deform, which slows down its release.

%\end{document}

\end{document}